**Some Comments on the Universal Constants.**

J. Dunning-Davies,
Department of Physics,
University of Hull,
England HU6 7RX.
J.Dunning-Davies@hull.ac.uk

**Abstract.**

Recent references to the commonly accepted expression for the entropy of a black hole to questions concerning the constancy of some of the so-called 'universal constants of nature' are questioned, as is the validity of the said entropy expression.

**Introduction.**

The notion that some of the commonly accepted 'universal constants' are not in fact constant has been around for quite a long time, certainly extending back to the work of such as Dirac [1] and Milne [2]. In more recent times, a varying speed of light has been advocated [3] and has been seen to explain some of the problems facing cosmology, eliminating the need for inflationary theories. However, the very recent articles linking possible constraints on the variation of these constants with the accepted theory of black holes are certainly open to question [4,5].

**Entropy and Universal Constants.**

According to Planck [6], 'The entropy of a physical system in a definite state depends solely on the probability of this state'. Based upon the statistical independence of independent events and the additivity of entropies of separate systems, this dependence is found to be logarithmic. Any constants in the argument of the logarithm appear as additive constants, and Boltzmann deliberately left an additive constant in the entropy undetermined, as is done in all of classical thermodynamics. The only universal constant to appear is that in the constant factor of proportionality. Boltzmann worked in moles, Planck in molecules, and so it was Planck who determined Boltzmann's constant, $k$ and this is the only 'universal' constant to appear naturally in the expression for entropy. Other universal constants creep into entropy expressions through the introduction of equations of state into the basic relation. A classic example of this is provided by black body radiation, where Planck's constant and the speed of light in vacuo appear in the Stefan-Boltzmann constant.

Heisenberg remarked [7] some years ago that, in order to introduce a mass, a fundamental length must be found for only then can a minimum mass be defined by its Compton wavelength. Once this is accomplished, the charge may be introduced via the 'classical' radius of the electron, $e^2/mc^2$. This introduction of a finite radius goes beyond quantum theory though since Planck's constant, $\hbar$, does not appear. The constant $\hbar$ is seen to separate the classical theory of heat from the quantum theory of black body radiation and the constant, $c$, separates Newtonian from relativistic mechanics but the constant $e$ has no such role. Hence, the electric charge or, equivalently, the size of the fine structure constant, $e^2/\hbar c$, must await explanation.

Conventional wisdom decrees that the entropy of a black hole is proportional to the area of its event horizon [8,9] and, for an uncharged, non-rotating black hole, the widely accepted expression is

$$S_{bh} = \frac{4\pi k G}{\hbar c} M^2,$$

where $k$ is Boltzmann's constant, $G$ the universal constant of gravitation, $\hbar$ Planck's constant, $c$ the speed of light and $M$ the mass of the black hole. It might be noted that this expression for the entropy of an uncharged, non-rotating black hole, the so-called Bekenstein-Hawking expression, shares a common feature with the entropy of black body radiation; it does not contain an arbitrary constant. In the case of black body radiation, this is vitally important since, if this were not so, the entropy would not tend to zero with temperature. However, from the equation above, it is clearly seen from the derivative that the temperature is inversely proportional to the mass of the black hole and, as a result, the entropy will tend to infinity as the temperature tends to zero, - in clear violation of Nernst's heat theorem! Again, it might be noted that, if the above entropy is parameterised in terms of the temperature, a decrease in the speed of light would result in a decrease in the entropy at constant temperature. This is contrary to what is claimed [4] if it is treated as a function of mass. The problem here is that the above black hole entropy expression is not a truly fundamental expression for the entropy, certainly not in the sense that that for the entropy of black body radiation is when expressed in terms of the internal energy and volume, since the equation of state introduces the Stefan-Boltzmann constant. Incidentally, it might be noted also that, as mentioned some years ago [10], if true, the above entropy expression for a black hole does not permit the use of several well-known thermodynamic expressions. More importantly, the result has been shown [11] to lead to violation of the Second Law of Thermodynamics. Hence, its validity must be open to question!

As has been pointed out previously [11], with the undoubted benefit of hindsight, it might be felt that Planck could have focussed beneficially on finding a fundamental relation for entropy which contained, in addition to $k$, one other fundamental constant - $\hbar$. Boltzmann's 'lottery', as Lorentz [12] liked to call it, always contained such a constant, although he paid no attention to its physical significance. To Boltzmann it was simply a mathematical trick enabling him to count discrete entities and, in any case, at the end of his calculation it was always allowed to tend to zero thereby taking the continuum limit. Planck, however, was allowed no such luxury but had to grapple with its physical meaning. He introduced two constants, $k$ and $\hbar$. The first distinguished the macroscopic from the microscopic; the second, the classical theory of heat from quantum theory. The constant $c$ appeared only in the classical calculation of the number of Planck oscillators in a finite frequency interval. These constants, together with the universal constant of gravitation $G$, could be used to construct units of mass, length, time, and temperature and Planck [13] speculated that they would 'retain their significance for all times and all cultures, including extraterrestrial and nonhuman ones.' Incidentally, Planck went on to comment that 'these *natural units* would retain their natural significance as long as the laws of gravitation and the propagation of light in vacuum and the two laws of thermodynamics retain their validity'. Hence, Planck seemed to feel that questioning universality and the fundamental constants tantamount to questioning the two most important laws of thermodynamics!

**Conclusion.**

The whole question of the constancy of the so-called 'universal constants of nature' has been around for a long time, as indicated by the early references cited here. However, work is still ongoing in this area. As far as the universal constant of gravitation is concerned, for example, measurements have been being made of it since Cavendish's attempt, based on a suggestion by Michell, in 1798 [14]. When all the measurements made over the intervening years are considered, the value of this universal constant of gravitation would seem to be increasing with time very slowly. However, the more recent, more accurate experiments seem to indicate that it is, in fact, constant in time, although there are suggestions that its value varies with position over the earth's surface. Again, as mentioned in the introduction, it has been suggested that the speed of light is not a constant but varies as the square root of the background temperature. If true, this would revolutionise much scientific thinking but it is, as yet really only a theoretical suggestion. This whole area is obviously one that requires a lot more investigation but the constancy, or otherwise, of the normally accepted constants of nature remains an open question.

**References.**